\documentclass[11pt]{article}
\usepackage{blois,epsfig}

\def\be{\begin{equation}}
\def\ee{\end{equation}}
\def\bea{\begin{eqnarray}}
\def\eea{\end{eqnarray}}

\begin{document}
\vspace*{4cm}
\title{Deconstructed Higgsless Models at LHC: The Top Triangle Moose \footnote{E.H. Simmons presented this work at the 2010 {\it Rencontres de Blois}; it summarizes work published as Ref. 6.}}

\author{R.Sekhar Chivukula, Neil D. Christensen, Baradhwaj Coleppa, Elizabeth H. Simmons}

\address{Department of Physics and Astronomy, Michigan State University\\ East Lansing, Michigan 48824, USA}

\maketitle\abstracts{
We describe a deconstructed Higgsless model in which electroweak symmetry breaking results largely from a Higgsless mechanism while the top quark mass is generated by a topcolor mechanism.  The top quark mass arises from a Yukawa coupling to an effective top-Higgs which develops a small vacuum expectation value.  Both the heavy partners of the electoweak gauge bosons and those for the SM fermions can be light enough to be visible at LHC.}

\section{Introduction}
 Deconstructed \cite{Deconstruction-Georgi,Deconstruction-Hill} Higgsless \cite{Csaki-Reference} models provide valuable insight into the mechanism of electroweak symmetry breaking (EWSB) without the presence of a scalar particle in the spectrum.  The ``bulk'' of the extra dimension is replaced by a chain of gauge groups strung together by non linear sigma model fields. The spectrum typically includes extra sets of charged and neutral vector bosons and heavy fermions. A general analysis of Higgsless models (see refs. in \cite{Chivukula:2009ck}) suggests that to satisfy precision electroweak constraints, the standard model (SM) fermions must be ``delocalized'' into the bulk.  A useful realization of this idea, ``ideal fermion delocalization'' \cite{IDF}, dictates that the light fermions be delocalized so as not to couple to the heavy charged gauge bosons. The simplest framework capturing these ideas is the ``three site Higgsless model''\cite{threesiteref}, with just one gauge group in the bulk and correspondingly, only one set of heavy vector bosons. The twin constraints of getting the correct value of the top quark mass and having an admissible $\rho$ parameter push the heavy fermion masses into the TeV regime in that model.

This presentation reviews Ref. 6, which decouples these constraints by separating the mechanisms that break the electroweak symmetry and generate the masses of the third family of fermions. Specifically, we modify  the three-site model by adding a ``top Higgs'' field, $\Phi,$ that couples preferentially to the top quark (see Figure 1). We thereby obtain a massive top quark and heavy fermions in the sub TeV region, without altering tree level electroweak predictions.

The idea of a top Higgs is motivated by top condensation models, and our model is most closely aligned with topcolor assisted technicolor theories first proposed in Ref. 7,  in which EWSB occurs via technicolor \cite{Eichten:1979ah,Dimopoulos:1979es} interactions while the top mass has a dynamical component arising from topcolor \cite{Hill-Topcolor-1,Hill-Topcolor-2} interactions and a small component generated by extended technicolor.   The dynamical bound state arising from topcolor dynamics can be identified as a composite top Higgs field, and the low-energy spectrum includes a top Higgs boson. The extra link in our ``top triangle moose'' model that corresponds to the top Higgs field results in the presence of uneaten Goldstone bosons, the top pions, which couple preferentially to the third generation. The model can thus be thought of as the deconstructed version of a topcolor assisted technicolor model.

\section{The Model}

\begin{figure}[t]
\begin{center}
\includegraphics[width=1.25in]{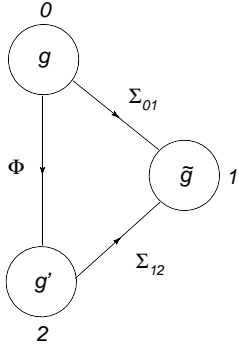}
\caption{The gauge structure of the ``top triangle moose'' model in Moose notation. The $SU(2)$ coupling $g$ and $U(1)$ coupling $g'$ of sites 0 and 2 are approximately the SM weak and hypercharge gauge couplings, while the much stronger $SU(2)$ coupling $\tilde{g}$ represents the 'bulk' gauge coupling.}
\label{fig:Triangle}
\end{center}
\end{figure}

We now introduce the features required to study the model's LHC phenomenology; more detail, including references to the related BESS \cite{BESS-1} and hidden local symmetry \cite{HLS-1} ideas, is in Ref.~\cite{Chivukula:2009ck}.  As shown in Moose notation in Figure \ref{fig:Triangle}, the extended electroweak gauge structure is $SU(2)_0\times SU(2)_1\times U(1)_2$.  The SM fermions derive their $SU(2)$ charges mostly from site 0 (which is most closely associated with the SM $SU(2)$) and the bulk fermions, mostly from site 1. 

The non linear sigma field $\Sigma_{01}$ breaks the $SU(2)_0\times SU(2)_1$ gauge symmetry down to $SU(2)$, and field $\Sigma_{12}$ breaks $SU(2)_1 \times U(1)_2$ down to $U(1)$. The left handed fermions are $SU(2)$ doublets residing at sites 0 ($\psi_{L0}$) and 1 ($\psi_{L1}$), while the right handed fermions are a doublet under $SU(2)_{1}$($\psi_{R1}$) and two $SU(2)$-singlet fermions at site 2 ($u_{R2}$ and $d_{R2}$). The fermions $\psi_{L0}$, $\psi_{L1}$, and $\psi_{R1}$ have SM-like $U(1)$ charges ($Y$):  $+1/6$ for quarks and $-1/2$ for leptons. Similarly, the fermion $u_{R2}$ ($d_{R2}$) has an SM-like $U(1)$ charge of $+2/3$ ($-1/3$); the right-handed leptons, likewise, have  $U(1)$ charges corresponding to their SM hypercharge values. The third component of isospin, $T_3$, takes values $+1/2$ for ``up'' type fermions and $-1/2$ for ``down'' type fermions, just like in the SM. The electric charges satisfy $Q=T_{3}+Y$.

We add a ``top-Higgs'' link to separate top quark mass generation from EWSB. The top quark couples preferentially to the top Higgs link via the Lagrangian:
\begin{equation}
\mathcal{L}_{top}=-\lambda_{t}\bar{\psi}_{L0}\,\Phi\, t_{R}+h.c.\label{top quark mass L}
\end{equation}
When $\Phi$ develops a non zero vacuum expectation value (vev), Eqn.(\ref{top quark mass L}) generates a top quark mass term. To ensure that most EWSB comes from the Higgsless side, we choose the vev's of $\Sigma_{01}$ and $\Sigma_{02}$ to be $F=\sqrt{2}\,v$ $\textrm{cos}\,\omega$  and the vev associated with the top Higgs sector to be $f=\langle\Phi\rangle=v$ $\textrm{sin\,}\omega$ (where $\omega$ is small). The top Higgs sector also includes uneaten Goldstone bosons: the top pions. We assume they are heavy enough not to affect electroweak phenomenology. 

Light fermion mass terms arise from fermion couplings to  the non linear sigma fields
\begin{eqnarray}
\mathcal{L} & = & M_{D}\left[\epsilon_{L}\bar{\psi}_{L0}\Sigma_{01}\psi_{R1}+\bar{\psi}_{R1}\psi_{L1}+\bar{\psi}_{L1}\Sigma_{12}\left(\begin{array}{cc}
\epsilon_{uR} & 0\\
0 & \epsilon_{dR}\end{array}\right)\left(\begin{array}{c}
u_{R2}\\
d_{R2}\end{array}\right)\right].
\label{eqn:Light fermion mass}
\end{eqnarray}
We denote the Dirac mass  setting the scale of the heavy fermion masses as $M_D$.  Here, $\epsilon_{L}$ is a flavor-universal parameter describing delocalization of the left handed fermions. All the flavor violation for the light fermions is encoded in the last term; the delocalization parameters for the right handed fermions, $\epsilon_{fR}$, can be adjusted to realize the masses and mixings of the up and down type fermions. For this phenomenological study, we assume that all the fermions, except the top, are massless and hence will set these $\epsilon_{fR}$ parameters to zero.  

The \emph{tree level} contributions to precision measurements in Higgsless models come from the coupling of standard model fermions to the heavy gauge bosons.  Choosing the profile of a light fermion bilinear along the Moose to be proportional to the profile of the light $W$ boson makes the fermion current's coupling to the $W'$ vanish because the $W$ and $W'$ fields are mutually orthogonal. This procedure (called ideal fermion delocalization \cite{IDF})  keeps deviations from the SM values of all electroweak quantities at a phenomenologically acceptable level.  We find that the ideal delocalization condition in this model is $\epsilon_{L}^{2}=M_W^2 / 2 M_{W'}^2$, as in the three-site model.  Likewise, the $W'$ phenomenology is the same as in the three-site model; and the projected reach of LHC $W'$ searches is summarized in Figure \ref{fig:Luminosity plot} .

The top quark mass matrix may be read from Eqns. (\ref{top quark mass L}) and (\ref{eqn:Light fermion mass}) and is given by:
\begin{equation}
\left(\begin{array}{cc}
M_{D}\epsilon_{tL} & \lambda_{t}v\textrm{sin}\omega\\
M_{D} & M_{D}\epsilon_{tR}\end{array}\right).
\label{top mass matrix}
\end{equation}
Diagonalizing the top quark mass matrix perturbatively in $\epsilon_{tL}$
and $\epsilon_{tR}$, we find:
\begin{equation}
m_{t}= \lambda_{t}v\,\textrm{sin}\,\omega\left[1+\frac{\epsilon_{tL}^{2}+\epsilon_{tR}^{2}+\frac{2}{a}\epsilon_{tL}\epsilon_{tR}}{2(-1+a^{2})}\right], \qquad\qquad a\equiv\frac{\lambda_{t}\, v\,\textrm{sin}\omega}{M_{D}},
\label{top mass}
\end{equation}
so that $m_{t}$ depends mainly on $v$ and only slightly on $\epsilon_{tR}$, in contrast to the three-site model, where $m_t \propto M_D \epsilon_L \epsilon_{tR}$.  Since the $b_L$ is the weak partner of the $t_L$, its delocalization is (for
$\epsilon_{bR}\simeq 0$) also determined by $\epsilon_{tL}$. Thus, the tree-level $Zb\bar{_{L}b_{L}}$coupling can constrain $\epsilon_{tL}$. We find $g_{L}^{Zbb}$ retains its tree-level SM value if the $t_L$ is  delocalized like the light fermions: $\epsilon_{tL} = \epsilon_L$.

Finally, the contribution of the heavy top-bottom doublet to $\Delta\rho$ is of the same form as in the three-site model \cite{threesiteref}: $\Delta\rho= M_{D}^{2}\,\epsilon_{tR}^{4} / 16\,\pi^{2}\, v^{2}$.
But since the top quark mass is dominated by the vev of the top Higgs instead of $M_{D}$, now $\epsilon_{tR}$ can be as small as the $\epsilon_R$ of any light fermion. There is no conflict between  a large top quark mass and a small  value of $\Delta\rho$. Thus, the heavy fermions in the top triangle moose can be light enough to be produced at LHC.

\section{Heavy quarks at the LHC}

\subsection{Pair production: $pp\rightarrow Q\bar{Q}\rightarrow WZqq\rightarrow lll\nu jj$} \label{Subsection:pair}

Pair production of heavy quarks occurs at LHC via gluon fusion and quark annihilation processes, with the former dominating for smaller $M_D$.  Each heavy quark decays to a vector boson and a light fermion. For $M_{D}<M_{W',Z'}$, the decay is purely to the standard model gauge bosons. We study the case where one heavy quark decays to $Z+j$ and the other decays to $W+j$, with the gauge bosons subsequently decaying leptonically. Thus, the final state is $lll \nu jj$.  In addition to particle identification cuts, we impose cuts to improve the signal-to-background ratio: both jets should be central and have high $p_T$, the dilepton invariant mass should be close to $M_Z$.  Imposing the full set of cuts described in Ref. 6 eliminates the SM background.   Figure \ref{fig:Luminosity plot} shows the integrated LHC luminosity required for a 5$\sigma$ discovery signal.  When $M_{D}\geq \textrm{900 GeV}$ and $M_{W'}\leq M_{D}$ there will not be enough signal events for the discovery of the heavy quark since the decay channel $Q \to W' q$ becomes significant. To explore this region, we now investigate the single production channel where the heavy quark decays to a heavy gauge boson.

\subsection{Single production: $pp\rightarrow Qq\rightarrow W'qq'\rightarrow WZqq'$} \label{Subsection:single}

\begin{figure}[tb]
\begin{center}
\includegraphics[width=3in]{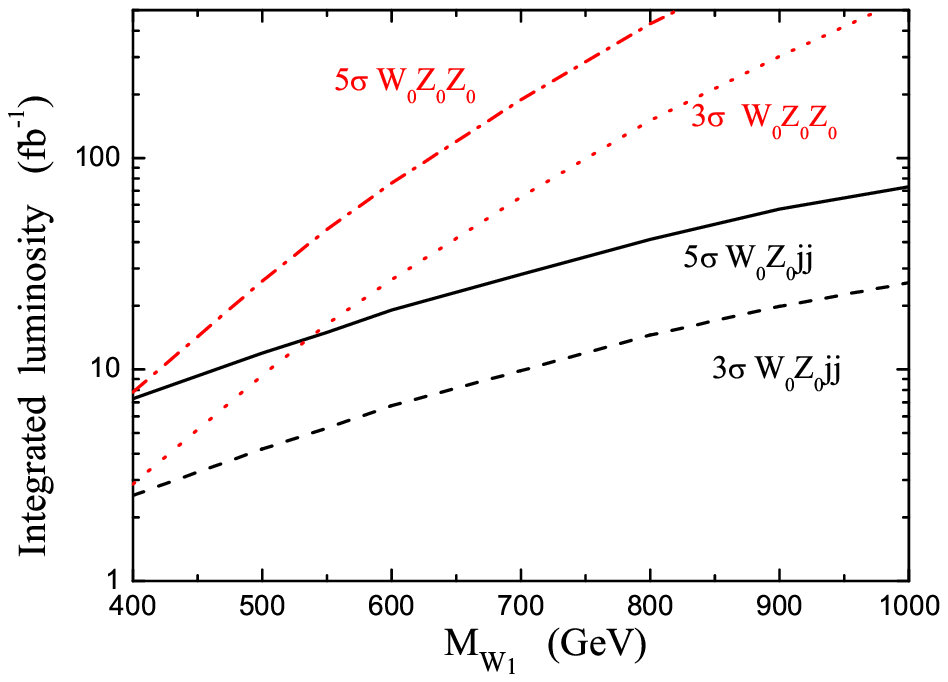} \hspace{1cm} \includegraphics[width=2.75in]{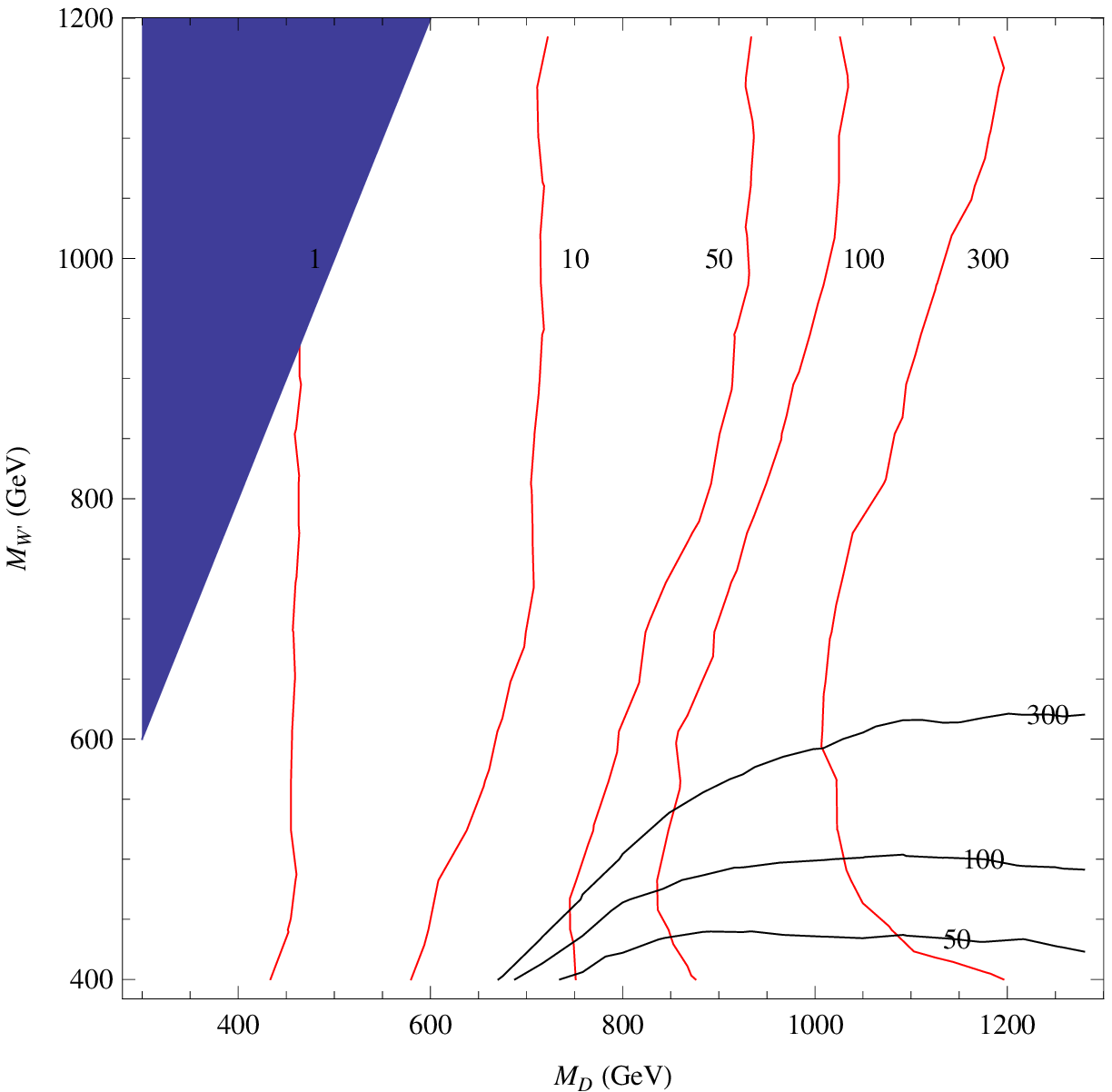}
\caption{Left pane: Integrated LHC luminosities required [Ref. 14] for $3\sigma$ and $5\sigma$ level $W'$ detection in the vector boson fusion (lower) and associated production (upper) channels. Right pane: Luminosity required [Ref. 6] for a $5\sigma$ discovery of the heavy vector fermions at the LHC in the single (horizontal curves) and pair (vertical curves) production channels for given values of the heavy fermion and $W'$ masses. The two channels are complementary.}
\label{fig:Luminosity plot}
\end{center}
\end{figure}

While the single production channel  is electroweak, the smaller cross section is compensated by the fact that the $u$ and $d$ are valence quarks, and their parton distribution functions fall less sharply than the gluon's.  Also, there is less phase space suppression in the single production channel than in the pair production case. We analyze the processes $\left[u,u\rightarrow u,U \right ]$, $\left [d,d\rightarrow d,D  \right ]$ and $\left [u,d\rightarrow u,D \ {\rm or}\ U,d\right ]$, which occur through a $t$ channel exchange of a $Z$ and $Z'$. Since we want to look at the region of parameter space where $M_{W'}$ is smaller than $M_{D},$ we let the heavy quark decay to a $W'$. The $W'$ decays 100\% of the time to  $W + Z$, because its coupling to two SM fermions is zero in the limit of ideal fermion delocalization. We constrain both the $Z$ and $W$ to decay leptonically so the final state is $lll \nu jj$. The signal-enhancing cuts must be modified for this channel (see Ref 6):  the jet from heavy quark decay will still be hard and central, but the other jet will be soft and forward.  As a result, the SM background is non-zero. Figure \ref{fig:Luminosity plot} shows the integrated luminosity required to achieve a 5$\sigma$ discovery signal in this channel.  Almost the entire parameter space is covered by the combination of pair and single production modes.

\section*{Acknowledgments}
This work was supported in part by US National Science Foundation grant PHY-0354226.

\end{document}